\documentclass[AMA,Times1COL]{WileyNJDv5} 

\articletype{Applied Research}%

\received{Date Month Year}
\revised{Date Month Year}
\accepted{Date Month Year}
\journal{Journal}
\volume{00}
\copyyear{2023}
\startpage{1}

\usepackage{amssymb}
\usepackage{lineno}

\newcommand{\be}{\begin{equation}}
\newcommand{\ee}{\end{equation}}
\newcommand{\ba}{\begin{array}}
\newcommand{\ea}{\end{array}}
\newcommand{\bx}{\begin{matrix}}
\newcommand{\ex}{\end{matrix}}

\usepackage{graphicx,times}
\usepackage{graphicx,epstopdf}
\usepackage{amsmath}
\usepackage{latexsym}
\usepackage{color}  
\usepackage[T1]{fontenc}
\usepackage[utf8]{inputenc}
\usepackage{usr1}
\usepackage{psfrag}
\usepackage{makecell} 
\usepackage{multicol}
\usepackage{multirow}
\usepackage{placeins}
\usepackage{subcaption}
\usepackage{epstopdf}
\usepackage{hyperref}
\usepackage{algpseudocode}
\usepackage{algorithm}
\usepackage[nameinlink]{cleveref}
\usepackage{svg}
\makeatletter
\def\tagform@#1{\maketag@@@{\ignorespaces#1\unskip\@@italiccorr}}
\let\orgtheequation\theequation
\def\theequation{(\orgtheequation)}
\makeatother
\DeclareMathOperator{\Aop}{\large{\mathsf{A}}}

\begin{document}

\title{Simulation of parametrized cardiac electrophysiology in three dimensions using physics-informed neural networks}

\author[1]{Roshan Antony Gomez}
\author[1]{Julien St{\"o}cker}
\author[1]{Bar\i\c{s} Cans\i z}
\author[1]{Michael Kaliske}

\authormark{Gomez \textsc{et al.}}
\titlemark{Prediction of parametrized cardiac electrophysiology in three dimensions using physics-informed neural networks}

\address[1]{\orgdiv{Institute for Structural Analysis}, \orgname{Technische Universit\"at Dresden, 01062 Dresden}, \orgaddress{\country{Germany}}}

\corres{Michael Kaliske \email{michael.kaliske@tu-dresden.de}}



\abstract[Abstract]{Physics-informed neural networks (PINNs) are extensively used to represent various physical systems across multiple scientific domains. The same can be said for cardiac electrophysiology, wherein fully-connected neural networks (FCNNs) have been employed to predict the evolution of an action potential in a 2D space following the two-parameter phenomenological Aliev-Panfilov (AP) model. In this paper, the training behaviour of PINNs is investigated to determine optimal hyperparameters to predict the electrophysiological activity of the myocardium in 3D according to the AP model, with the inclusion of boundary and material parameters. An FCNN architecture is employed with the governing partial differential equations in their strong form, which are scaled consistently with normalization of network inputs. The finite element (FE) method is used to generate training data for the network. Numerical examples with varying spatial dimensions and parameterizations are generated using the trained models. The network predicted fields for both the action potential and the recovery variable are compared with the respective FE simulations. Network losses are weighed with individual scalar values. Their effect on training and prediction is studied to arrive at a method of controlling losses during training.}



\maketitle


\renewcommand\thefootnote{\fnsymbol{footnote}}
\setcounter{footnote}{1}

\section{Introduction}\label{sec:Intro}
The heart, being one of the most vital organs in the human body, is a very important area of scientific research towards the diagnosis and treatment of diseases. Current scientific progress is mainly dependent on clinical trials on humans and animals, which, apart from being ethically questionable \cite{Frank2023Ethics}, are also expensive. Furthermore, animal trials may not always be extrapolatable to human beings as described by Garcia et al.\cite{Garcia_2017:Challenges}. Research in computational cardiology aims to alleviate this dependency by providing numerical alternatives. Intricacies in the form of complex geometry and material composition and high degrees of heterogeneity and anisotropy are the principle challenges faced by such models. Biophysical models such as the finite element (FE) method \cite{Cansiz_2017:Computational} and data driven models are the primary approaches found in literature. For instance, Cansiz\cite{Cansiz_2017:Computational} applies the FE method to simulate fibrillation and subsequent defibrillation via cardiac resynchronization therapy in a personalized biventricle model generated from magnetic resonance imaging. Niederer et al.\cite{niedererCreationApplicationVirtual2020} gives a comprehensive review of numerical models in computational cardiology. 

While biophysical models are capable of producing high fidelity results, their use is restricted in real-time and parameterized scenarios due to high computational overhead. Patient-specific model parameters necessitate unique modeling and running separate simulations. On the other end of the spectrum, data driven approaches are rapid, at the expense of solution accuracy. Reduced order models (ROMs) \cite{chellappaFastReliableReducedorder2023a} are the next logical step from full order biophysical models to reduce computational time, but they tend to be inefficient for tasks such as prediction of cardiac electrophysiology (EP), where phenomena like wave re-entries fail to be accurately captured. An excellent attempt at balancing the trade-off between these models is seen in \cite{frescaPODEnhancedDeepLearningBased2021}, where deep learning (DL) based ROM is used to solve cardiac EP tasks. The training is done using snapshots from full order model (FOM) solutions after applying randomized proper orthogonal decomposition (POD). Although remarkable success is achieved, the model continues to rely on a discretized geometry, restricting model performance to the level of discretization. In this context, physics-informed neural networks (PINNs) show potential to simulate cardiac EP phenomena without reliance on neither the discretization type nor density, all the while maintaining coherence to the governing physical laws. They can be seen a step towards trade-off balance from fully data driven models.

PINNs are an extension of traditional artificial neural networks (ANNs) to include the governing physics as an additional constraint to the network in the form of additional losses, thereby reducing the requirement of training or ground truth (GT) data. ANNs have been categorized as universal approximators by Hornik et al.\cite{HORNIK1989359_universal} They have been used extensively in scientific domains such as weather forecasting \cite{zenkner2023forecast}, structural design optimization \cite{Mote_2019_designopt}, disease diagnosis and progress identification \cite{azeem2021_diseasedetect} and many more. Their success may be attributed to the availability of large amounts of GT points. This creates a bottleneck in the case of using ANNs for simulating the behavior of the human heart, as it is difficult to collect accurately measured data for the individual primary variables of the governing equations. For example, techniques for the measurement of the action potential remain an area of active research \cite{Hayes2019APmeasure}. Additionally, data driven models are known not to meet the solution accuracy standards set by traditional biophysical models. 

Hurtado et al. \cite{costabal2020} and Grandits et al.\cite{grandits2021} have used PINNs to estimate the arrival times and the conduction velocity of the action potential at high resolution in the left atrium in sinus rhythm conditions by solving the eikonal equation. The model may be used to generate high resolution activation time maps from low resolution clinical measurements. The eikonal equation is a simple diffusion equation relating activation times to conduction velocity and is insufficient for the study of cardiac arrhythmia. Martin et al.\cite{martin2021eppinns} have used PINNs to successfully predict the propagation of the action potential at the two-dimensional (2D) level under healthy as well as arrhythmia-induced conditions, by solving the diffusion equation with the Aliev-Panfilov\cite{ALIEV1996293} (AP) model in the monodomain setting. Data from finite difference (FD) simulations is used to train the PINN and a homogeneous isotropic material is considered. Being an excellent initial study, it advocates strongly for PINNs as a solid foundation to build on towards an accurate and time-efficient solution algorithm for cardiac EP. While it motivates research in generalizing the model to higher dimensions and more anisotropic conditions, it is expected that the demand on computational resources for such an application would be too high. 

This study aims to generalize the use of PINNs to 3D parameterized EP tasks while minimizing computational overhead, by comprehensively studying the training behavior of the neural network, to discover possible obstacles and formulate mitigation strategies. Such an angle of approach is unexplored in \cite{martin2021eppinns}. Taking inspiration, the work done by Martin et al.\cite{martin2021eppinns} is extended by training PINNs for the prediction of cardiac EP in 3D using normalized inputs and consistently scaled PDEs, both of which are unaccounted for in \cite{martin2021eppinns}. Loss weights are employed, which are scalars multiplied with each individual network loss. The effect of the selection of loss weights on the learning process of the network is thoroughly investigated, as it is known to help to improve training by maintaining a balance between the loss terms as explained by Noakoasteen et al.\cite{PINN_EM_2020}. Moreover, a novel staged reduction of loss weights is implemented, which is seen to improve both training stability and prediction accuracy. Therewith, a scroll wave is simulated in a three-dimensional cuboid of cardiac tissue under isotropic conditions using the AP model\cite{ALIEV1996293}. The network is further trained by including the timing of application of an external stimulus as an additional input, so that its effect on the generation and sustenance of scroll waves may be studied.

This paper is organized as follows.
\Cref{sec:CardiacEP} introduces the AP equations \cite{ALIEV1996293}, which provide the source term and an internal variable evolution equation. Further, the governing equation is described in its strong form, considering spatial electrical conductivity. In \Cref{sec:PINNs}, the concept of PINNs is introduced in the context of cardiac EP. The two stages involved in the learning process, namely the forward and backward passes, are explained along with a system for the consistent scaling of the PINN residuals. Likewise, methods used to control the training behavior, such as gradient clipping and a staged reduction of loss weights are also introduced. \Cref{sec:NumericalEx} illustrates the investigative numerical examples, starting with the PINN prediction of the action potential at the cellular level, with parameterization of material and boundary quantities. Here, the learning behavior of the PINNs to predict both continuous and discontinuous solutions are compared. The second example sees the PINNs being trained to predict scroll waves within a cuboidal domain, with a staged reduction of loss weights being applied to reduce prediction errors. In the last example, the PINN parameterizes the time of application of external stimuli, to show its applicability to conduct sensitivity studies. \Cref{sec:FEM} describes the weak form of the governing equation, its FE discretization, and a temporal discretization via an implicit backward Euler scheme. Furthermore, the equation is consistently linearized to be solved using the Newton method and a local Newton iteration is employed to calculate the evolution of the internal variable.

\section{Cardiac electrophysiology}\label{sec:CardiacEP}
      
    Kumar et al.\cite{kumar2009robbins} explains that the human heart is a very complex electromechanical system that works continuously throughout the life span of an individual, pumping around 6000 litres of blood per day, which is regulated by its electrical conduction system. Its primary components include the sinoatrial (SA) node, the atrioventricular (AV) node, the conduction axes (Bundle of His, Purkinje fibres) as well as the cardiac muscle. The contraction and relaxation of the cardiac muscle occur concurrently with a surge in electric potential (depolarization) followed by a gradual decrease to baseline (repolarization).
    Gopal et al.\cite{GOPAL2014611cardiacEP} describes cardiac EP as a sub-speciality of cardiology dealing with heart rhythm disorders. Studying the electrical activity within the heart is crucial in this field. Most of the research is done with the help of electrocardiogram measurements, as laid out in detail by Roberts\cite{12LeadEKG}. Several mathematical models have been formulated that are able to accurately mimic the depolarization characteristics of the cardiac muscle cells, which may be broadly classified as ionic and phenomenological models. Ionic models take the concentrations of ions within the cardiac cells and the ionic current through voltage-controlled gates on the cell membrane into consideration. These models are used to study intricate phenomena such as the fundamental mechanisms underlying initiation and sustenance of arrhythmia in the heart as done by Bishop and Plank\cite{Bishop2012TheRO}. 
    
     Phenomenological models essentially capture the effects of ionic currents through parameters, which when calibrated using experimental data, can be used to mimic mechanisms with a lower, yet acceptable level of accuracy. The FitzHugh-Nagumo model \cite{FITZHUGH1961445,Nagumo_1962} and the AP model are exemplars. The AP model \cite{ALIEV1996293} with the diffusion equation provides the physics information for the neural network. The weak form of the model is also used to generate FE simulations, which play the role of both training as well as testing data for the PINNs in later sections. The electrophysiological behavior of non-oscillatory cells such as the ventricle wall muscles are governed by the AP equations, written as
     
    \begin{subequations}\label{Eq:f_phi}
    \begin{gather}
         f^\phi = c\phi\left[\phi-\alpha\right]\left[1-\phi\right]-r\phi+I \text{,}\\
         f^r = \left[\gamma + \frac{\mu_1r}{\mu_2 + \phi}\right]\left[-r-c\phi\left[\phi-b-1\right]\right]\text{,}
    \end{gather}
    \end{subequations}

    with \(\alpha\), \(a\), \(b\), \(c\), \(\mu_1\), \(\mu_2\) and \(\gamma\) being material parameters. The recovery variable \(r\) models the restitution of the action potential to base level and \(I\) describes the externally applied stimuli.
        
    It is to be noted that the derivatives are taken with respect to a normalized time \(\tau\), which, along with \(\phi\), are defined by
    \be\label{Eq:norm}
    \phi = \frac{\Phi+\delta_\phi}{\beta_\phi} \quad \text{and} \quad \tau=\frac{t}{\beta_t}\text{.}
    \ee
    Here, \(\Phi\) and \(t\) are the action potential and time in physical units, namely mV and ms, while \(\phi\) and \(t\) are the respective non-dimensional normalized values as seen in the original FitzHugh-Nagumo and AP models. The normalization scalars (\(\delta_\phi\), \(\beta_\phi\), \(\beta_t\)) for the two models are given in \autoref{tab:NormAP}. 
\begin{table*}[h]
\centering
\caption{Normalization scalars used in models for cardiac electro-physiology} 
\label{tab:NormAP}
\begin{tabular}{ccc}
  Parameter             & FitzHugh-Nagumo model   & Aliev-Panfilov model   \\ [0.5ex]
  \hline                             \\ [-1.5ex]
  $\beta_\phi$ [mV]     & 65                      & 100                    \\ [0.5ex]
  $\delta_\phi$ [mV]    & 35                      & 80                     \\ [0.5ex]
  $\beta_t$ [ms]        & 220                     & 12.9                   \\ [0.5ex]
  \hline  
\end{tabular}

\end{table*} 

    The solution of the AP model from the cellular perspective (without considering spatial propagation) by discretization along the time domain using the Newton method for solving non-linear equations is shown in \autoref{Fig:Ali_cell}.
    
    \inputfig[htpb]{figures/sec2/Graphs}{diff-cont}
    
    In this study, the myocardial domain \(\cal B\) is assumed to be an isotropic, homogeneous continuum with boundary \(\delta\cal B\), which is further split into the essential boundary \(\delta\cal B_\phi\) and natural boundary \(\delta {\cal B}_{q}\). A mono-domain model is assumed as seen in \cite{Cansiz_2017:Computational} and \cite{goktepe09novelfem} which, unlike the bi-domain model, does not discriminate between intra- and extracellular media. The governing equation in its strong form can be written as

    \begin{equation}\label{Eq:AP_strong}
        \dot{\Phi}=\text{div}(\boldsymbol{q})\;+\;F^\Phi\text{,}
    \end{equation}
    with the Dirichlet and Neumann boundary conditions, respectively, reading 
    \begin{subequations}\label{Eq:AP_bounds}
    \begin{gather}
         \Phi = \Tilde{\Phi} \quad \text{on} \quad \delta\cal B_\phi \quad \text{and}\\
         \boldsymbol{q}\cdot\boldsymbol{n}=\Tilde{q} \quad \text{on} \quad \delta {\cal B}_{q}\text{,}
    \end{gather}
    \end{subequations}
    with \(\boldsymbol{n}\) being the unit outward normal vector on \(\delta\mathcal{B}_q\). Additioanlly, the initial condition is written as 
    \be
        \Phi = \Phi_0 \quad \text{at} \quad t=0 \text{.}
    \ee
    Furthermore, the vector representing the flux of the action potential is written as 
    \be\label{Eq:Flux}
    \boldsymbol{q}=\boldsymbol{D}\cdot\nabla\Phi\text{.}
    \ee

    Considering \(d_\text{iso}\) and \(d_\text{ani}\) as the isotropic and anisotropic conductivities, respectively, the conductivity tensor \(\boldsymbol{D}\) may be written as an additive decomposition of an isotropic part (\(\boldsymbol{D}_\text{iso}=d_\text{iso}\boldsymbol{I}\)) and an anisotropic component (\(\boldsymbol{D}_\text{ani}=d_\text{ani}\boldsymbol{f_0}\otimes\boldsymbol{f_0}\)) along the structural tensor (\(\boldsymbol{f_0} \otimes \boldsymbol{f_0}\))
    \be\label{Eq:Cond_decomp}
        \boldsymbol{D} = \boldsymbol{D}_\text{iso} + \boldsymbol{D}_\text{ani} \text{.}
    \ee
    The unit-corrected source term \(F^\Phi\) is defined by 
    \be\label{Eq:SourceConvert}
    F^\Phi = \frac{\beta_\phi}{\beta_t}f^\phi\text{,}
    \ee
    and the equation describing the evolution of the recovery variable is written as
        \be\label{Eq:drdt}
        \frac{\partial r}{\partial \tau}=f^r\text{.}
        \ee
\section{Physics-informed neural networks for cardiac \\
electrophysiology}\label{sec:PINNs}
PINNs have recently gained popularity as an algorithm for solving partial differential equations, as they are in theory able to capture the "physics" involved with a lower number of training points compared to traditional ANNs. ANNs are a class of machine learning algorithms that approximate functions between two vector spaces as a composition of weighted linear operators \(\left(\boldsymbol{T}\right)\) and non-linear scalar activation functions \(\left(\chi\right)\). The name originates from the observation that such a function can typically be expressed as a network of nodes and layers that resemble the architecture of neural connections, as depicted in the red box in \autoref{Fig:PINN_scheme}. A PINN is constructed from the basic ANN model by adding a layer after the output to calculate the multi-objective loss function \(\left(\mathcal{L}\right)\) as a sum of the residuals (\(R_i\)) and boundary conditions (\(\text{BC}_i\)) that represent the governing PDEs and boundary conditions. This section gives a brief description of PINNs as proposed by Lagaris et al.\cite{Lagaris_1998}, and later re-introduced by Raissi et al.\cite{RAISSI2019686}. The PDEs under consideration in \autoref{Eq:AP_strong} are also modified to account for the normalization of the input space, which is common practice for neural networks to improve training convergence as explained by Santurkar et al.\cite{batchnorm2019}
    \subsection{The forward pass}\label{subsec:ForwardPass}
    Each hidden layer \(l\) consists of \(n_l\) nodes, and is associated with a second order weight tensor \(\boldsymbol{W}_l\) represented as a matrix  of size \(n_{l} \times n_{l-1}\), where \(l=1, 2, ..., L\) and a constant bias vector $\boldsymbol{\mathsf{b}_l}$ of size $n_l$. For instance, the weight matrix of the first layer would be of size  \(n_{1} \times n_{0}\), where \(n_0\) is the number of dimensions of the input space. The output of each layer is calculated as

    \be\label{Eq:ANN_layerOP}
    \boldsymbol{v}_l=\chi\left(\boldsymbol{T}_l\left(\boldsymbol{v}_{l-1}\right)\right)\text{,}
    \ee
    
    where
    
    \be\label{Eq:ANN_linearOP}
    \boldsymbol{T}_l\left(\boldsymbol{v}_{l-1}\right)=\boldsymbol{W}_l\cdot\boldsymbol{v}_{l-1}+\boldsymbol{\mathsf{b}_l}\text{.}
    \ee
    
    The activation function \(\chi\left(\boldsymbol{x}\right)\) accounts for the possible non-linearities in the approximated function.
    \inputfig[htbp]{figures/sec4/Diagrams}{diff-cont}
     
    As established by Hornik et al.\cite{HORNIK1989359_universal}, if a deterministic relationship exists between a set of inputs and outputs, a multi-layer feed-forward network can approximate any measurable function to any desired degree of accuracy, given the use of a sufficient number of hidden layers, neurons and the appropriate activation function. Consider a GT point \(\boldsymbol{\bar x_\text{GT}}=\left[ \bar x, \bar y, \bar z, \bar t \right]^\top\) in the input space of real numbers \(\mathbb{R}^4\) at which the solution \(\mathcal{\boldsymbol{p}}(\boldsymbol{\bar x_\text{GT}})=\left[\phi, r\right]^\top\) in the solution space \(\mathbb{R}^2\) is known. The network generates an output \(\mathcal{\boldsymbol{\hat{p}}}(\boldsymbol{\bar x_\text{GT}})=\left[\hat{\phi}, \hat{r}\right]^\top\) during the first of the two parts of the algorithm called the forward pass. The map \(\boldsymbol{M}\) from the input vector \(\boldsymbol{\bar x_\text{GT}}\) to the output \(\mathcal{\boldsymbol{\hat{p}}}(\boldsymbol{\bar x_\text{GT}})\) is defined as

    \be\label{Eq:ANN_forwardPass}
    \boldsymbol{M}:\mathbb{R}^{4} \mapsto \mathbb{R}^{2}: \boldsymbol{\bar x_\text{GT}} \mapsto \left(\boldsymbol{T}_L \circ\: \chi_{L-1} \circ ... \circ\: \chi_{1} \circ\: \boldsymbol{T}_1 \right) (\boldsymbol{\bar x_\text{GT}}) = \mathcal{\boldsymbol{\hat{p}}}(\boldsymbol{\bar x_\text{GT}})\text{.}
    \ee

    The error between the values predicted by the network and the ground truth \((\mathcal{\boldsymbol{p}}(\boldsymbol{\bar x_\text{GT}}))\) is used to define the loss of the network, the calculation of which marks the end of the forward pass, and the start of the backward pass. Since the task at hand is a regression problem, a mean-squared-error (MSE) loss function

    \be\label{Eq:ANN_loss}
    \mathcal{L_\text{GT}} = \frac{1}{m}\sum_{i=1}^{m} (\mathcal{\boldsymbol{\hat{p}_i}}(\boldsymbol{\bar x_\text{GT}})-\mathcal{\boldsymbol{p_i}}(\boldsymbol{\bar x_\text{GT}}))^2
    \ee

    is chosen, where the sum is taken across \(m\) training points in the input domain. The choice of MSE as the loss function is heavily swayed by its convex nature with a definite global minimum, which facilitates the convergence of the gradient descent algorithm used during the backward pass. Further details regarding the backward pass are explained in \hyperref[subsec:BackwardPass]{Subsection~\ref*{subsec:BackwardPass}}.
    
    For PINNs, the loss function also acts as a means to pass the physical information in the PDEs to the network. An additional layer is added in between the output layer and the final loss as seen in \autoref{Fig:PINN_scheme}, at which the PDE and boundary condition residuals are calculated using automatic differentiation (AD) as explained by Baydin et al.\cite{baydin2018_AD} AD calculates the derivatives using a static or dynamic computational graph from the inputs to the outputs, and is used by most modern machine learning frameworks. For a collocation point \(\boldsymbol{\bar x_\text{c}}=\left[ \bar x, \bar y, \bar z, \bar t \right]^\top\) in \(\mathbb{R}^4\) at which the solution is unknown, the network predicts the solution \(\mathcal{\boldsymbol{\hat{p}}}(\boldsymbol{\bar x_\text{c}})\) using the above defined map \(\boldsymbol{M}\). With a differential operator \(\mathsf{D[\cdot]}\) and a boundary operator \(\mathsf{B[\cdot]}\), a typical system of PDEs can be written as

        \be
            \mathsf{D}[\mathcal{\boldsymbol{p}}] = 0\text{,}\quad\mathsf{B}[\mathcal{\boldsymbol{p}}] = 0.
        \ee

    From the predicted approximation \(\mathcal{\boldsymbol{\hat{p}}}(\boldsymbol{\bar x_\text{c}})\), the residuals \(\mathsf{D}[\mathcal{\boldsymbol{\hat{p}}}]=[R_1,\;R_2,\;...,\;R_{n_\text{PDE}}]^\top\) and \(\mathsf{B}[\mathcal{\boldsymbol{\hat{p}}}]=[\text{BC}_1,\;\text{BC}_2,\;...,\; \text{BC}_{n_\text{BC}}]^\top\) are calculated using AD. Due to the random initialization of network weights \(\boldsymbol{W_0}\), even with well established weight initialization methods such as the one proposed by Glorot and Bengio\cite{xavier-glorot2010}, the PDE losses can be uncontrollably high, resulting in poor learning for the PINN. To compensate this issue, it is beneficial to multiply each individual loss with a coefficient \(\alpha\). With the loss weights considered, the MSE loss functions \(\mathcal{L}_{Rj}\) and \(\mathcal{L}_{\text{BC}k}\) for the \(j^\text{th}\) PDE and the \(k^\text{th}\) boundary condition and \(n\) collocation points read

    \begin{subequations}\label{Eq:PINN_weighted_loss_MSE}
    \begin{gather}
        \mathcal{L}_{Rj} = \frac{\alpha_j}{n}\sum_{i=1}^{n} R_j(\mathcal{\boldsymbol{\hat{p}_i}}(\boldsymbol{\bar x_\text{c}}))^2\text{,} \\
        \mathcal{L}_{\text{BC}k} = \frac{\alpha_k}{n}\sum_{i=1}^{n} \text{BC}_k(\mathcal{\boldsymbol{\hat{p}_i}}(\boldsymbol{\bar x_\text{c}}))^2\text{.}
    \end{gather}
    \end{subequations}
    The final loss for \(n_\text{PDE}\) PDEs and \(n_\text{BC}\) boundary conditions with \(m\) GT points and \(n\) collocation points are written as

    \be\label{Eq:PINN_weighted_loss}
    \mathcal{L} = \sum_{j=1}^{n_\text{PDE}}{\mathcal{L}_{Rj}} + \sum_{k=1}^{n_\text{BC}}{\mathcal{L}_{BCk}} + \mathcal{L_\text{GT}}\text{.}
    \ee
    
    \subsection{Consistent scaling of residuals}\label{subsec:PINNRes}
    According to Bishop\cite{bishop1995neural}, it is nearly always advantageous to apply pre-processing transformations to the input data before it is presented to a network, which has not been considered in \cite{martin2021eppinns}. Intuitively, normalization of inputs would ensure that \(\boldsymbol{T}_l\) would stay within a finite range (usually \([0,1]\) or \([-1,1]\)) so that the derivatives of the non-linear function \(\chi\) do not vanish. A mathematical treatment of the reasons behind the effectiveness of input normalization has been written by Santurkar et al.\cite{batchnorm2019} 
    
    When inputs are to be normalized for PINNs, the corresponding PDEs are also required to be consistently scaled. In this paper, all the input variables are normalized to an interval \([a^*, b^*]\). If \(x_i \in [p^*_i, q^*_i]\) is an input variable in its original domain, it can be normalized to \(\Bar{x}_i \in [a^*, b^*]\) using the relationship

    \be\label{Eq:i/p_norm}
    \Bar{x}_i = a^* + \frac{(b^*-a^*)(x_i-p^*_i)}{(q^*_i-p^*_i)}\text{.}
    \ee
    For the task at hand, \({p_i}\) and \(q_i\) are the limits for the input variables \(x_i\), such that \(x_1=x,\; x_2=y,\; x_3=z\) and \(x_4 = t\), with \(\boldsymbol{x}=[x,\;y,\;z,\;t]\) being normalized to \(\Bar{\boldsymbol{x}}=[\Bar{x},\;\Bar{y},\;\Bar{z},\;\Bar{t}]^\top\).
    
    Owing to the use of non-linear activation functions \(\chi\) such as the hyperbolic tangent function, the output \(\mathcal{\boldsymbol{\hat{p}}}(\boldsymbol{x})\) naturally lies in the range \([-1,1]\). This can be taken advantageously by expressing \autoref{Eq:AP_strong} with respect to the normalized action potential \(\hat{\phi}\), and a scaled value \(\hat{r}=0.4r\) of the recovery variable. Applying the relationships in \hyperref[Eq:norm]{Equations~\ref*{Eq:norm}} and \ref{Eq:SourceConvert} and using the concept of the total derivative and the chain rule, the PDEs in \autoref{Eq:AP_strong} may be written in residual form as 

    \begin{subequations}\label{Eq:AP_strong_normed}
    \begin{gather}
        R_1 = \frac{\text{d}\hat{\phi}}{\text{d}\Bar{t}} - (b^*-a^*)(q^*_4 - p^*_4)\Bar{C} - \frac{(q^*_4-p^*_4)f^\phi}{(b^*-a^*)\beta_t}\text{,} \\
        R_2 = \frac{\text{d}\hat{r}}{\text{d}\Bar{t}} - 0.4\frac{(q^*_4-p^*_4)f^r}{(b^*-a^*)\beta_t}\text{,}
    \end{gather}
    \end{subequations}

    where

    \begin{equation}\label{Eq:C_bar}
        \begin{split}
        \Bar{C} = \sum_{i=1}^3\frac{D_{ii}}{(q^*_i-p^*_i)^2}\frac{\partial^2\hat{\phi}}{\partial \Bar{x_i}^2} &+
            \frac{D_{12}+D_{21}}{(q^*_1-p^*_1)(q^*_2-p^*_2)}\frac{\partial^2\hat{\phi}}{\partial \Bar{x_1}\Bar{x_2}} \\
            &+\frac{D_{23}+D_{32}}{(q^*_2-p^*_2)(q^*_3-p^*_3)}\frac{\partial^2\hat{\phi}}{\partial \Bar{x_2}\Bar{x_3}} + 
            \frac{D_{31}+D_{13}}{(q^*_3-p^*_3)(q^*_1-p^*_1)}\frac{\partial^2\hat{\phi}}{\partial \Bar{x_3}\Bar{x_1}}
                \text{,}
        \end{split}
    \end{equation}
    
    with \(D_{ij}\) representing the components of the conductivity tensor \(\boldsymbol{D}\), for \(i, j = 1,2,3\). Similar to the PDEs, Dirichlet and Neumann boundary conditions can also be represented as residuals. For example, the boundary conditions yield

    \begin{subequations}\label{Eq:BC_conditions}
    \begin{gather}
        \text{BC}_1 = \hat{\phi}-\phi_\text{GT} \quad \text{for} \: \boldsymbol{\Bar{x}} \in \{\boldsymbol{\Bar{x}}_\text{GT}\} \text{,}\\
        \text{BC}_2 = \frac{\partial\hat{\phi}}{\partial\boldsymbol{\Bar{x}}}=\boldsymbol{0}\quad\text{for} \: \boldsymbol{\Bar{x}} \in \partial\mathcal{B}\text{.}
    \end{gather}
    \end{subequations}  
    
    \subsection{The backward pass}\label{subsec:BackwardPass}
    The backward pass is an optimization task, where the objective function \(\mathcal{L}\) is to be minimized with respect to the weights \(\boldsymbol{W}\) of the network. Among the various algorithms available for optimization, the first-order optimizer Adam developed by Kingma et al.\cite{kingma2017adam} is employed here, which builds on the gradient descent algorithm (GDA). 
    
    The GDA starts with an initial guess of the optimum weight tensor \(\boldsymbol{W_0}\) to complete one iteration of the forward pass, estimating the initial loss \(\mathcal{L}\). The backpropagation algorithm (BPA) introduced by Werbos\cite{Werbos} is utilized to calculate its gradient with respect to the weights \(\boldsymbol{W}\). As shown by Kreyszig et al.\cite{kreyszig11}, the gradient of a scalar-valued function with respect to its input vector points in the direction of its steepest ascent. The weights are then updated as 

    \be\label{Eq:GDA_weight_update}
    \boldsymbol{W} = \boldsymbol{W_0} - \alpha_\text{NN}\frac{\partial\mathcal{L}}{\partial\boldsymbol{W_0}} \text{,}
    \ee
    where \(\alpha_\text{NN}\) is called the learning rate. The combined forward pass followed by the backward pass through all of the available training examples is called an epoch. The algorithm tends to converge to a point \(\boldsymbol{W_\text{opt}}\) such that

    \be\label{Eq:1st_order_opt}
    \frac{\partial\mathcal{L}}{\partial\boldsymbol{W_\text{opt}}} \approx \boldsymbol{0}\text{.}
    \ee
    
    If second-order approaches like the Newton algorithm is used for the optimization task, they take the curvature of the objective function into account along with the direction of steepest descent  \cite{Optalgos2023}. The weights are updated at each epoch according to

    \be\label{Eq:2nd_order_weight_update}
    \boldsymbol{W} = \boldsymbol{W_0} - {\mathcal{L}^{\prime\prime}(\boldsymbol{W_0})}^{-1}\mathcal{L}^\prime(\boldsymbol{W_0}) \text{,}
    \ee
    in which \(\mathcal{L}^{\prime\prime}\) represents the second derivative of the loss with respect to the weights. This increase in the required order of derivatives results in the generation of excessive computational overhead, in addition to that which inherently exists for PINNs in the form of the additional layer of PDE and boundary losses as seen in \autoref{Fig:PINN_scheme} and explained in \hyperref[subsec:ForwardPass]{Subsection~\ref*{subsec:ForwardPass}}. 
    
    To this end, the Adam optimizer is used, which is one the most popular optimization algorithms in machine learning. As summarized by Kingma et al.\cite{kingma2017adam}, the algorithm calculates individual adaptive learning rates for different parameters from estimates of the exponential moving averages of the gradient \(m_e\) and the squared gradient \(v_e\) of the loss function \(\mathcal{L}\) at epoch \(e\). Now, \autoref{Eq:GDA_weight_update} is modified to

    \be\label{Eq:Adam_weight_update}
    \boldsymbol{W} = \boldsymbol{W_0} - \alpha_\text{NN}\frac{m_e}{\sqrt{v_e}} \text{.}
    \ee
    It is noted that both \(m_e\) and \(v_e\) are functions of the gradient of the loss \(\mathcal{L}\) at the current epoch \(e\), and other hyperparameters which control the decay rates of the learning rate.  
    
    \subsection{Gradient clipping}\label{subsec:GradClip}
    There have been several improvements made to the base PINN as seen in \cite{RAISSI2019686}, to tackle various issues related to its training when applied to tasks with high complexity which comes in the form of higher order derivatives, complex geometries, uneven solution landscapes and discontinuities. Gradient surgery by Yu et al.\cite{yu2020gradientsurgery}, learning rate annealing and an improved network architecture by Wang et al.\cite{wang2021gradpath}, utilization of weak form residuals and importance sampling by Nabian et al.\cite{Nabian_2021_ImportanceSampling} implemented in the package NVIDIA Modulus as seen in \cite{hennigh2020nvidia} are some effective examples to name a few. 
    
    A positive, yet essential side effect to the application of such treatments is the reduction in exploding gradients, which is a common problem while enforcing PDEs in their strong form with higher dimensionality. The work by Wang et al.\cite{wang2021gradpath} depicts a detailed analysis of gradient pathologies in PINNs. In the paper at hand, exploding gradients have been tackled by a much simpler solution called gradient clipping. The interested reader is referred to Zhang et al.\cite{zhang2020gradientclip}, where it is shown that first order optimization with gradient clipping converges much faster than gradient descent with a fixed step size, by analyzing gradient smoothness. For every epoch, the Euclidean norm \(l_g\) of the gradient vector \(\boldsymbol{g}\) is scaled down to a predefined value \(l_c\) called the clipping norm, before applying in \hyperref[Eq:GDA_weight_update]{Equations~\ref*{Eq:GDA_weight_update}} or \ref{Eq:Adam_weight_update}. This is done intuitively in an attempt to avoid drastic steps along the surface of the objective function 

    \be\label{Eq:GradClip}
    \boldsymbol{g} = \displaystyle\frac{\boldsymbol{g_0}}{l_g}l_c \text{,}
    \ee
    where \(\boldsymbol{g_0}\) is the vectorized form of the tensor \(\displaystyle \alpha_\text{NN}\frac{\partial\mathcal{L}}{\partial\boldsymbol{W_0}}\).
\section{Numerical Examples}\label{sec:NumericalEx}
    The solution of the AP model using PINNs is attempted in stages, starting from the cellular level as seen in \hyperref[Eq:f_phi]{Equation~\ref*{Eq:f_phi}}. In addition to the normalized time \(\Bar{t}\), a normalized material parameter \(\Bar{c}\) is also included as an input variable to the network, to study the ability of the network to generate parameterized solutions. The material parameter \(c\) controls the time interval between depolarization and repolarization, which changes inversely with the change in \(c\). The effect of gradient clipping on the network performance is also studied. The input dimensions are then increased to include all three spatial dimensions as in \autoref{Eq:AP_strong}, without the additional material parameter. Moreover, the timing of application of external stimuli is included as an additional input to the network to generate a parameterized solution, as a step in the direction of being able to conduct numerical studies of the myocardium under arrhythmic conditions in the future. In all the examples presented below, only Dirichlet boundary conditions have been imposed in the form of GT data from the FE models. Neumann boundary conditions have been avoided to reduce computational complexity as much as possible, as it is found from a previous study that the requirement of Neumann boundary conditions to calculate higher order derivatives of the output variables through AD increases the instability of the training process. Further details regarding the hyperparameters used during the experiment can be found in \autoref{tab:PINN_HypParam_example_1}.   

\begin{table*}[h]
\centering
\caption{Hyperparameters used for Example 1} 
\label{tab:PINN_HypParam_example_1}
\begin{tabular}{m{3.5cm}|c|c}
  Hyperparameter                            &    Continuous  & Discontinuous               \\
  \hline                             
  Boundary conditions                       & $\text{BC}_\text{GT}$  & $\text{BC}_\text{GT}, \text{BC}_1$  \\
  Normalized limits $[a^*,b^*]$           & [0,1]          & [0,1]                       \\ 
  Learning rate $(\alpha_{NN})$             & 0.005          & 0.005                       \\ 
  Collocation points                        & 200000         & 200000                      \\ 
  Testing points                            & 10000          & 10000                       \\ 
  Optimizer                                 & Adam           & Adam                        \\ 
  Epochs                                    & 100000         & 100000                      \\ 
  Clip norm $(l_c)$                         & 0.1            & 0.1                         \\ 
  \hline  
\end{tabular}
\end{table*} 


    \subsection{Network architecture and parameters}\label{subsec:PINN_parameters}
    For all the examples shown in this paper, PINNs have been implemented using DeepXDE, a Python framework by Lu et al.\cite{lu2021deepxde} The inspiration for the initial architecture and parameter use is drawn from the strategies adopted by Martin et al.\cite{martin2021eppinns}
    It consists of \(L=5\) layers of \(n_l=60\) nodes each and the hyperbolic tangent function \(tanh(x)\) is used as activation function, except for Example 1 in \hyperref[subsec:ex1_time_para]{Subsection~\ref*{subsec:ex1_time_para}}, where \(L=3\) layers of \(n_l=50\) nodes each is found to be sufficient. The normalization constants used for every example have been summarized in \autoref{tab:PINN_HypParam_example_1}, and the material parameters can be seen below in \autoref{tab:Mat_Params}.

\begin{table*}[h]
\centering
\caption{Simulation parameters and boundary conditions used for the numerical examples} 
\label{tab:Mat_Params}
\begin{tabular}{c|ccc}
  Parameter    & Example 1     & Example 2     & Example 3         \\ 
  \hline                                                                     
  $\alpha$ $[-]$           & 0.05          & 0.01          & 0.05              \\ 
  $\gamma$ $[-]$           & 0.002         & 0.002         & 0.002             \\ 
  $b$ $[-]$                & 0.15          & 0.15          & 0.15              \\ 
  $c$ $[-]$                & [4,8]         & 8             & 8                 \\ 
  $\mu_1$ $[-]$            & 0.2           & 0.2           & 0.2               \\ 
  $\mu_2$ $[-]$            & 0.3           & 0.3           & 0.3               \\ 
  $d_\text{iso}$ $[\frac{\text{mm}^2}{\text{ms}}]$ & -             & 1.2           & 1.2               \\ 
  $I$ $[-]$                & 10            & 5             & 5                 \\ 
  $\tau_\text{stim}$ $[-]/t_\text{stim}$ $[\text{ms}]$ & 30            & 490            & [410,530]                \\ 
  \hline  
\end{tabular}\;
\begin{tabular}{c|c}
  Boundary condition   & Loss description \\ [0.5ex]
  \hline
  $\text{BC}_\text{GT}$          & $\hat{\phi} - \phi_\text{GT} = 0$ \\ [0.5ex]
  $\text{BC}_1$          & $\hat{\phi} - 1 = 0\:\text{when}\:\tau=0$ \\ [0.5ex]
  $\text{BC}_2$          & $\hat{\phi} = 0\:\text{when}\:\tau=\tau_\text{max}$ \\ [0.5ex]
  \hline
\end{tabular}
\end{table*} 
    
    \subsection{Example 1: Prediction at the cellular level}\label{subsec:ex1_time_para}
    The phrase "cellular level" refers to the change of \(\phi\) and \(r\) with time at a point \(\boldsymbol{x}\) in the spatial domain as described by \hyperref[Eq:f_phi]{Equation~\ref*{Eq:f_phi}} and seen in \autoref{Fig:Ali_cell}. A fraction of the points generated by the Newton iteration strategy explained in \Cref{sec:FEM} is used as GT points for training and evaluation. The inputs to the network are normalized according to \autoref{Eq:i/p_norm}. The residuals take the form

    \begin{subequations}\label{Eq:ResEx1}
    \begin{gather}
        R_1 = 0.01\frac{\text{d}\hat{\phi}}{\text{d}\Bar{t}} - [\Bar{c}(q_2-p_2)+p_2)]\hat{\phi}(\hat{\phi}-\alpha)(1-\Bar{\phi})+2.5\hat{r}\hat{\phi} - I\text{,}\\
        R_2 = 0.025\frac{\text{d}\hat{r}}{\text{d}\Bar{t}} - \left(\gamma + \frac{2.5\mu_1\hat{r}}{\mu_2+\hat{\phi}}\right)(-2.5\hat{r}-[\Bar{c}(q_2-p_2)+p_2]\hat{\phi}(\hat{\phi}-b-1))\text{.}\\
    \end{gather}
    \end{subequations}  
    Two sets of simulations are generated by the Newton iteration with the value of \(c\) varying in the range \([3, 9.5]\), with initial conditions \(\phi=1\) at \(\tau=0\), \(I=0\) and  \(\phi=0\) at \(\tau=0\), \(I=5\) at \(\tau=\tau_\text{stim}\) respectively. The former set of boundary conditions yields the solution to be continuous in time by virtue of the non-existence of \(I\), as opposed to the latter set as seen in \autoref{Fig:Ali_cell}. This is expected to minimize complexity for the PINN, while maintaining consistency with \hyperref[Eq:f_phi]{Equation~\ref*{Eq:f_phi}}. Other material parameters chosen are given in \autoref{tab:Mat_Params}. The best combination of hyperparameters found for the network are then used to find the same for learning the discontinuous function. The hyperparameters for the Adam optimizer have been set to default according to Kingma et al.\cite{kingma2017adam} The root-mean-squared error (RMSE) \(e\) between the prediction \(\hat{\phi}\) by the PINNs and the solution \(\phi\) by the Newton method is calculated as

    \be\label{Eq:RMSE}
    \displaystyle e = \sqrt{\frac{\sum_{i=1}^n (\phi_i - \hat{\phi_i})^2}{n}} \text{,}
    \ee
    which is used to evaluate the performance of the network. Here \(n\) is the total number of points generated by the Newton method.   

    \subsubsection{Continuous solution}\label{subsec:AP_time_cont}
    The last term \(I\) of the residual \(R_1\) is detrimental to the learning of the PINN, as it introduces a so-called "stiffness" \cite{sharmaStiffPDEsPhysicsInformedNeural2023} in the PDE due to the discontinuity at \(\tau=\tau_\text{stim}\). The training is done by increasing the number of GT points shown to the network each time. A total of 200000 collocation points is uniformly sampled in the input domain, and 10000 points are used for testing. 
  
    It can be seen that gradient clipping significantly improves the prediction capability of the network, the effect of which can also be seen in the loss plots during training for both cases in \autoref{Fig:Loss_comp}. A total of 51000 points is generated by the Newton method, from which GT points are sampled randomly for training. The predictions of the network are shown in \autoref{Fig:Result_ex1_cont} along with the exact solution for different sample sizes of GT points.
    
    \inputfig[htb]{figures/sec5/ex1-Loss-cont}{diff-cont}
    With the PINN, a final MSE of \(2.32 \times 10^{-2}\) is achieved with 1.5 \% of the total GT points, which captures a rough shape of the solution curve as seen in the first column of \autoref{Fig:Result_ex1_cont}. The effect of increasing the material parameter \(c\) can also be seen qualitatively in the prediction. An increase in the percentage of GT points to 10 \% is necessary to achieve an MSE of \(3.6 \times 10^{-5}\), the same order of which is achieved with a mere 0.3 \% GT points with gradient clipping as seen in columns 2 and 3 of \autoref{Fig:Result_ex1_cont}. The advantage of using PINNs over conventional ANNs is also evident from the accurate prediction of the recovery variable, despite the absence of GT points for \(r\). It is also noteworthy that the PINN predicts the curve with a larger accuracy at higher values of \(c\), the reason for which is a topic of ongoing research but is out of the scope of this study.
    \inputfig[htbp]{figures/sec5/ex1-cont}{diff-cont}
    
    \subsubsection{Discontinuous solution and time parameterization}\label{subsec:AP_time_discont}
    A discontinuity is introduced in the \(\phi\)-curve of the solution by setting
    \be\label{Eq:ExtStim_1}
    I = 
    \begin{cases}
        10 & \tau \in [\tau_\text{stim}-\epsilon, \tau_\text{stim}+\epsilon] \\
        0  & \text{otherwise} \\
    \end{cases}\text{,}
    \ee
    during the generation of GT points. Here, \(\epsilon\) is a tolerance set to the order of $10^{-6}$. This represents an external stimulus, which can be used to initiate a depolarizing wave front to simulate scroll waves as shown by Cansiz\cite{Cansiz_2017:Computational}. The timing \(\tau_\text{stim}\) of the application of the external stimulus determines the generation and duration of scroll waves in the myocardium, which is suggestive of arrhythmia. From a simulation standpoint, finding the critical \(\tau_\text{stim}\) requires multiple runs of the simulation through trial and error. Here, an opportunity develops to take advantage of the parameterization capabilities of neural networks. 
    
    In this example, the effect of simply introducing \(I\) as a discontinuous function as seen in \autoref{Fig:Stim_ex1} is examined at first, followed by the addition of \(\tau_\text{stim}\) as an input to the network, in order to arrive at a parameterized model. The increase in complexity of the solution further increases the ruggedness of the objective function landscape, inciting a larger tendency for gradient explosions as seen in the loss plot in \autoref{Fig:LossStim_discont}.

    A comparison between the PINN predictions and GT points for the solution of the AP model with external stimuli and \(c\) as the parameterized quantity can be seen in \autoref{Fig:Result_ex1_discont_c}, which compares the effect of changes in loss weights \(\alpha_i\) on network predictions. An MSE of the order \(10^{-5}\) is achieved for \(\hat{\phi}\) with 1025 GT points. The effectiveness of the model may be better judged by its prediction of the internal variable \(\hat{r}\), for which no GT points are fed to the network during training. It is evident from \autoref{Fig:Result_ex1_discont_c} that the choice of values for \(\alpha_i\) influences the network's adherence to various PDEs. The first row of results shows that despite the success of the network in predicting values of \(\hat{\phi}\), it fails to accomplish proportional success for the prediction of \(\hat{r}\). In such scenarios, the network behaves more like a traditional neural network, and can be seen overfitting to the GT points. The loss weights \(\alpha_i\) are therefore adjusted accordingly to ensure adequate adherence to \autoref{Eq:f_phi} while avoiding convergence to a trivial solution. This also highlights the importance of including \(\hat{r}\) as a metric for evaluating the performance of the trained neural network.
    
    \inputfig[htb]{figures/sec5/ex1-discont}{diff-cont}

    Parameterization of \(\tau_\text{stim}\) has also been achieved with the same network architecture and 1025 GT points, resulting an MSE of the order of \(10^{-5}\). Due to the non-differentiability of \(I\), it is replaced by an exponential function

    \be\label{Eq:ExtStim_expo}
    I = 10 e^{\displaystyle{\left(-\frac{4.1}{2\epsilon}(\tau - \tau_\text{stim})\right)}^2}
    \ee
     as seen in \autoref{Fig:Stim_ex1}. The choice of the exponential function is inspired by the fact that \(tanh(x)\) is used as the activation function at all the nodes. It is to be noted that the width \(2\epsilon = 2 \times 10^{-6}\) TU of the excitation function is negligible compared to the magnitude of the time domain (100 TU) for the task at hand, and, therefore, justifies difference in shape between the square wave and the exponential function. The corresponding comparison between the PINN prediction and GT points is provided in \autoref{Fig:Result_ex1_timepara}. A minimum RMSE of \(1.1 \times 10^{-2}\) has been achieved by both of the trained models for the internal variable \(\hat{r}\). 

     \clearpage

    \inputfig[htbp]{figures/sec5/ex1-Loss-stim}{diff-cont}
    
    \subsection{Example 2: Prediction of 3D electrophysiology}\label{subsec:EP_3D}
    The number of input variables is further increased to include spatial conduction of electricity within the myocardium. The PDEs as seen in \autoref{Eq:AP_strong} are used with the scalar components of the conductivity tensor defined by \(d_\text{iso}=1.2\;\text{mm}^2/\text{ms}\) and \(d_\text{ani}=0\). The domain under consideration is a cubical piece of cardiac tissue of size \([100 \times 100 \times 100]\) mm. The boundary condition
    \be\label{Eq:IC_Ex2}
    \Phi = 
    \begin{cases}
        -40\;\text{mV} & x=0\;\text{mm},\:t=0\;\text{ms} \\
        -80\;\text{mV} & x\neq0\;\text{mm},\;t=0\;\text{ms} \\
    \end{cases}
    \ee
    
    is applied, initiating a depolarizing wave of action potential from \(x=0\;\text{mm}\) travelling along the x-axis with a wavefront parallel to the y-axis. Chaotic electrical activity within the myocardium is one among many signs of arrhythmia, which can be triggered by re-entrant waves such as scroll waves. These can be simulated by applying an external stimulus at tail end of the repolarization wave, as shown by Cansiz\cite{Cansiz_2017:Computational}. This is done by a square wave akin to \autoref{Fig:Stim_ex1}, except \(\epsilon\) being 10 ms, and the non-normalized time scale in ms being used, instead of \(\tau\).

    For the generation of GT points, the cube is discretized into 31 \(\times\) 31 \(\times\) 10 (9610), eight-noded hexahedral elements. The task is formulated according to \Cref{sec:FEM} with material parameters defined in \autoref{tab:Mat_Params}, and the time domain is discretized with a time step of \(\Delta t = 2\;\text{ms}\) in the interval \(t \in [0, 2000]\;\text{ms}\). The simulation generates a total of around \(1.1 \times 10^6\) GT points, out of which 11293 (0.1 \%) points are randomly sampled to be used as training data. 
    
    The PDEs are scaled according to \autoref{Eq:i/p_norm} with \(a^*=-1\) and \(b^*=1\) for all input variables. From the multiple trials conducted to assess the effect of the different training parameters, it is revealed that the relative magnitudes of the loss coefficients \(\alpha_i\) are of prime importance in aiding the network to learn efficiently. Despite scaling the PDEs, the increase in dimensionality exponentially increases the requirement for training data, resulting in a decrease in learning stability for the network \cite{Hughes_dimcurse1968}. Wang et al.\cite{wang2021gradpath} recommends the use of individual weights for each loss term, and annealing them according to moving averages of previous values of loss weights as training progresses, which is an approach similar to the annealing of the global learning rate by Kingma et al.\cite{kingma2017adam} The technique uses unit weights initially, which, for the task at hand, produces uncontrollably large initial PDE losses, eventually leading to gradient explosions, even though gradient clipping is employed. 

\begin{table*}[htbp]
\centering
\caption{Hyperparameters used in Example 2} 
\label{tab:Ex2-summ}
\begin{tabular}{l|c|c}
  Hyperparameter                            &    Trial 1     & Trial 2                     \\
  \hline                             
  Boundary conditions                       & $\text{BC}_\text{GT}$  & $\text{BC}_\text{GT}$ \\
  Normalized limits $[a^*,b^*]$           & [-1,1]         & [-1,1]                      \\ 
  Learning rate $(\alpha_{NN})$             & 0.005, 0.001   & 0.005, 0.001                \\ 
  Collocation points                        & 200000         & 200000                      \\ 
  Testing points                            & 10000          & 10000                       \\ 
  Optimizer                                 & Adam           & Adam                        \\ 
  Total epochs                              & 200000         & 200000                      \\
  Epochs (\(1^{\text{st}}\) stage)          & -              & 10000                       \\
  Epochs per stage                          & -              & 20000                       \\
  Initial loss weights $[\alpha_1,\;\alpha_2,\;\alpha_\text{GT}]$ 
                                            & $[10^{-5},\;10^{-9},\;1]$
                                            & $[10^{-5},\;10^{-9},\;1]$                    \\
  Loss weight decay rate \(\beta_l\)   & -              & $10^{-5}  $                 \\
  Clip norm $(l_c)$                         & 1.0            & 1.0                         \\ 
  \hline  
\end{tabular}
\end{table*} 
    
    A much more manual approach is taken in this paper, which involves finding via hyper-parameter studies, a combination of loss weights for the PDEs that allow the network to learn with some degree of stability. Moreover, the training is split into multiple batches of epochs, during which the loss weights are linearly adjusted. The rate and direction of adjustment are decided by observing the predictions of both the action potential \(\hat{\Phi}\) and the internal variable \(\hat{r}\) and comparing them to the solution generated by the FE method. As done in \hyperref[subsec:ex1_time_para]{Subsection~\ref*{subsec:ex1_time_para}}, the RMSE between the network outputs and the FE solution are also calculated, and 3D contours of the absolute error between them are plotted. It is also found that in addition to the global learning rate annealing done by the Adam optimizer used throughout the study, a manual step-wise decay of the learning rate \(\alpha_\text{NN}\) is beneficial for learning.

    \subsubsection{Trial 1: Base condition}\label{Ex2_tr1}

    The parameters used to successfully train the network in Example 1 are directly translated to a 3D setting. It is discovered from Example 1 and several other trials that, when the values of the loss weights \(\alpha_i\) are selected such that the initial value for each individual loss is smaller than \(10^{-1}\) and the total initial loss is less than 1, the network is able to train with sufficient stability.

     The initial losses are stored from 10 trial runs, and the average order is calculated. The loss weights are then set for the actual training run such that the conditions stated above are met.The loss weights \([\alpha_1,\;\alpha_2,\;\alpha_\text{GT}]=[10^{-5},\;10^{-9},\;1]\) are used, and the learning rate is decreased from 0.005 to 0.001 after 20000 epochs, for better convergence. A total of 160000 collocation points are sampled, out of which 100000 points are uniformly sampled within the input domain and 20000 points each at \(\Bar{x}=0,\;\Bar{t}=0\), \(\Bar{x} \ne 0,\;\Bar{t}=0\) and at \(\Bar{t}=\Bar{t}_\text{stim}\). The trained model is capable of estimating the action potential \(\hat{\Phi}\) with an RMSE of 7.13 mV. The contours of \(\hat{\Phi}\) in the first column are qualitatively similar to those of \(\Phi\) from the FE solution, as seen in \autoref{Fig:Result_ex2_T22}. An RMSE of \(0.54\) is obtained between \(\hat{r}\) and \(r\), which signifies a large deviation from ground truth when compared to Example 1. 
     
     \subsubsection{Trial 2: Staged reduction of loss weights}\label{Ex2_tr2}
     
     Tracking of losses during training of the network in trial 1 reveals that the high magnitude of GT loss prevents the network from learning the PDE losses. A lower initial value for \(\alpha_1\) leads to exploding gradients despite the use of clipping. The introduction of a periodic scaling down of the loss weight \(\alpha_1\) for \(R_1\) is introduced

    \be\label{Eq:stg_loss_wt}
    \alpha_1 = \frac{\alpha_0}{\beta_l \cdot e_f},
    \ee

     which enables the network to learn in the initial stages with a heavy reliance on GT data, followed by a greater emphasis on PDE residuals during the later stages of training. Other parameters are kept unchanged from trial 1. Here, \(\alpha_0\) is the initial loss weight for \(R_1\), \(\beta_l\) is the rate of weight decay and \(\displaystyle{e_f = \lfloor{e/20000}\rfloor}\) with \(e\) as the current epoch. In this example, the value of \(\alpha_1\) is updated every 20000 epochs. An improvement in fit is seen for both \(\Phi\) and \(r\), with RMSE values of 4.36 mV and \(0.17\), respectively.  Moreover, doubling the number of GT points used for training decreases the RMSE with \(\Phi\) to 2.68 mV and that with \(r\) to \(0.15\). A comparative plot of training loss between trials 1 and 2 can be seen in \autoref{Fig:Staged_loss_wt}, along with the change in loss weight $\alpha_1$. Moreover, a sensitivity analysis of the prediction error of the network with respect to the availability of interior GT points is seen in \autoref{Fig:Sample_study}. The respective contour plots generated by the trained models from trials 1 and 2 are shown in \hyperref[Fig:Result_ex2_T22]{Figures~\ref*{Fig:Result_ex2_T22}} and \ref{Fig:Result_ex2_T35}.          
    \inputfig[htb]{figures/sec5/ex2-Graphs}{diff-cont}
    
    \inputfig[htb]{figures/sec5/ex2-SampleStudy}{diff-cont}
    
    \inputfig[htb]{figures/sec5/ex2-T22}{diff-cont}
    
    \inputfig[htb]{figures/sec5/ex2-T35}{diff-cont}
    \clearpage 
    \subsection[p]{Example 3: Parameterized 3D electrophysiology}\label{subsec:EP_3D_param}

    As mentioned in \hyperref[subsec:AP_time_discont]{Subsection~\ref*{subsec:AP_time_discont}}, the generation of scroll waves within the myocardium is indicative of arrhythmia. The timing \(t_\text{stim}\) of the introduction of the external stimulus \(I\) is important to simulate a sustained scroll wave. The introduction of \(t_\text{stim}\) as an additional input to the PINN to train the network would enable the network to generate solutions for different values of \(t_\text{stim}\), saving time which would otherwise be required for multiple simulations. 

    For the generation of GT points, multiple FE simulations are run with \(t_\text{stim}\) varying from 400 ms to 520 ms at 30 ms intervals. From a total of \(56.3 \times 10^6\) GT points generated, 56320 (0.1 \%) points are used for training the network. A total number of 150000 collocation points are sampled uniformly within the domain, with an additional 80000 points at \({t}=0\;\text{ms}\). The 3D contour plot of \(\hat{\Phi}\) generated by the network for three distinct values of \({t}_\text{stim}\) is shown in \autoref{Fig:Result_ex3}.   

    \inputfig[htbp]{figures/sec5/ex3}{diff-cont}

    An RMSE of 5.6 mV is achieved between \(\hat{\Phi}\) and \(\Phi\). It can be seen that a scroll wave is generated and sustained when \(I\) is applied at the tail end of the depolarization wave at \(t_\text{stim}=480\;\text{ms}\). An earlier stimulus at \(t_\text{stim}=450\;\text{ms}\) generates a scroll wavefront which eventually vanishes by about \(t=800\;\text{ms}\), and one at \(t_\text{stim}=430\;\text{ms}\) does not cause any depolarization whatsoever.

\pagebreak
\clearpage
\section{Conclusions and outlook}\label{sec:Conclusion}
In this study, the use of PINNs to model the electrophysiological behavior of the myocardium is generalized from the study by Martin et al.\cite{martin2021eppinns} to a 3D setting. A fully connected NN model is employed, and the training process is optimized by observing changes in the training loss and prediction errors. The input variables are normalized with a corresponding modification of the residual equations for PDE loss, loss weights are applied and the effect of changing them dynamically during training is also explored. All the examples are trained on one RTX TITAN GPU, requiring a maximum training time of 3 hours.

Three numerical examples are described, each progressively increasing in dimensionality of the input domain. A novel control strategy for training the network is proposed, which involves a staged reduction of loss weights. Moreover, the network is trained with the parameters of the governing equations as additional inputs. It is seen that the examples generate solutions with RMSE values in the order of \(1\) mV, which translates to \(10^{-2}\) non-dimensional units. The order of error in the 3D setting is comparable to the implementation by Martin et al.\cite{martin2021eppinns} for the 2D case, but with only 11293 GT points as opposed to \(1.4 \times 10^5\) GT points which is a significant reduction. 

It is evident from the results that exploding gradients are the primary deterrents to learning, even in the most basic formulation of the task, and that gradient clipping is a simple solution to the issue, as seen in \autoref{Fig:Loss_comp}. From Example 1, it is observed that the use and proportioning of loss weights \(\alpha_i\) heavily influence the adherence of the trained model to the physics of the modelled system. The discrepancy between the errors in the prediction of \(\phi\) and \(r\) shows that it is important to examine all the output variables concerning the set of PDEs when evaluating the performance of the model. The inclusion of additional spatial dimensions and the electrical conductivity of the myocardium increases the order of the governing PDEs to 2, correspondingly increasing the complexity of the objective function landscape as well as prediction error. It was observed that a high magnitude of GT loss prevented the network from learning from PDE residuals and a reduced initial loss weight \(\alpha_1\) results in a higher probability of exploding gradients. As a novel mitigation strategy, a staged reduction of \(\alpha_1\) is applied to the prediction of scroll waves in a cubical slice of cardiac tissue. This allows the network to initially learn from GT data, and "take off the training wheels" in the later learning stages, to learn primarily from PDE residuals. This leads to a reduction in error accompanied with an increase in training stability. This method may be applied individually to different loss terms with distinct decay rates \(\beta_l\) to control the degree to which physics is informed. This would become necessary for a coupled system, such as an electromechanical model of the heart if it is to be modelled with multi-layered perceptrons using the PDEs in their strong forms. Training of the network to create parameterized models for material and boundary parameters such as \(c\) and \(t_\text{stim}\) is successful and causes significantly lower training instability compared to when the input layer is extended to include more primary variables. 

The results of the work done in this paper show that the use of PINNs for predicting the electrophysiological behavior of the myocardium in the three-dimensional domain requires balancing the PDE and boundary losses both initially and during training. A scroll wave phenomenon is successfully predicted, which is a typical precursor to chaotic electrical behavior, an indicator of cardiac arrhythmia. The model is capable of generating solution contours for different values of \(t_\text{stim}\) and captures its effect on the generation and sustenance of scroll waves. 

Although the electrical behavior can be modelled qualitatively, the prediction accuracy of the trained models is still too low for practical implementation. A first step towards the mitigation of prediction errors is the estimation of confidence intervals as seen in \cite{confidence-interval-96}. Furthermore, the current model is restricted in terms of the geometry it can accommodate, and it is important to incorporate complex geometries in order to eventually model the geometry of a real human heart. Methods such as \(\Delta\text{-PINNs}\) by Costabal et al.\cite{costabal2022deltapinns} can be an initial step into that direction. 
\pagebreak
\clearpage
\section{Materials and methods}\label{sec:MatMeth}

\subsection{Ethical statement}
None.






\appendix
\bmsection{FE implementation}\label{sec:FEM}
    In this section, the weak form of the governing equation as seen in \autoref{Eq:AP_strong} is derived and employed to generate ground truth data, which would then be used for both the training and validation of the PINNs. A Galerkin-type FEM according to Dal et al.\cite{dal2012} is used, and the FE stiffness is formulated as worked out by Cansiz\cite{Cansiz_2017:Computational}. The time integration along the discretized time domain is done using an implicit backward Euler integration scheme. The solution is obtained through an in-house version of FEAP (FE Analysis Program) \cite{taylor_feap_2014}. Since an external electric field is not required to be applied, the use of the mono-domain model presented in \autoref{Eq:AP_strong} is justified, according to Potse et al.\cite{Potse2006ACO}. 
    \subsection{Newton iteration for action potential }\label{subsec:Globaliter_phi}
        The weak form of the governing equation is obtained by multiplying the strong form in \autoref{Eq:AP_strong} by the square integrable weight function \(\delta \Phi\) such that \(\delta \Phi=0\) on \(\delta\cal B_\phi\). Integrating over the domain \(\cal B\) after applying integration by parts and the Gauss divergence theorem leads to
        \be\label{Eq:Weakphi}
        G^{\Phi}=\int_{\cal B} \delta\Phi\;\dot{\Phi}\; \text{d}V + \int_{\cal B} \nabla\delta\Phi\cdot\boldsymbol{q}\; \text{d}V
    				- \int_{\partial {\cal B}_q} \delta\Phi\; \Tilde{q} \;\text{d}A - \int_{\cal B} \delta\Phi\; F^{\Phi}\; \text{d}V = 0 \text{,}
        \ee
        with \(\Tilde{q}=\boldsymbol{q}\cdot\boldsymbol{n}\), where \(\boldsymbol{n}\) is the outward unit normal vector on \(\delta \cal B\).
        
        The domain is discretized by \(n_\text{el}\) elements \({\cal B}^e\), where \(e = 1, 2,...,n_\text{el}\). According to the isoparametric concept, the field variable \(\Phi\) and the weight function \(\delta\Phi\) are approximated by \(\Phi^h\) and \(\delta\Phi^h\), respectively, by interpolating between the $n_\text{n}$ nodes of each element using Lagrange shape functions \(N_i\), reading
        \be\label{Eq:Isoparam}
        \delta\Phi^h = \sum_{i=1}^{n_\text{n}} N_i\delta\Phi_i\text{,}\quad\Phi^h = \sum_{i = 1}^{n_\text{n}} N_i\Phi_i \text{.}
        \ee
        Temporal discretization is done by considering a time interval \(\left[t_n,t\right]\) with time increment \(\Delta t=t-t_n\), such that 
        \be\label{Eq:TimeDiscrete}
        \dot{\Phi}\approx\frac{\Phi-\Phi_n}{\Delta t} \text{.}
        \ee
        Here, \(\Phi\) and \(\Phi_n\) are the action potentials at \(t\) and \(t_n\), respectively. Substituting \hyperref[Eq:Isoparam]{Equations~\ref*{Eq:Isoparam}} and \ref{Eq:TimeDiscrete} into \autoref{Eq:Weakphi} and considering \(i\) and \(j\) as the node numbers ranging from \(1\) to \(n_n\) and \(I\) and \(J\) as the global degree of freedom numbers ranging from \(1\) to \(n_\text{dof}\), the discretized residual in the weak form may be written as
        \be\label{Eq:Weakphi_discrete}
        R^{\Phi}_I = \Aop^{n_\text{el}}_{e=1} \left[\int_{{\cal B}^e}\left(N_i\frac{\Phi-\Phi_n}{\Delta t} + \nabla N_i\cdot\boldsymbol{q}\right)\;\text{d}V
    						 - \int_{\partial{{\cal B}^e_q}} N_i\Tilde{q}\;\text{d}A
    						 - \int_{B^e} N_i F^{\Phi}\;\text{d}V \right]\text{.}
        \ee
        Here, \(\Aop\) represents the global assembly of contributions from each element in accordance with the element connectivity matrix. The weak form is linearized by increments \(\Delta\Phi\) of the action potential. The corresponding increment of the residual is found by calculating its directional derivative at each degree of freedom \(I\) with respect to \(\Phi_J\) at every other degree of freedom \(J\) along \(\Delta\Phi_J\), which is expressed as 
        \be\label{Eq:ResGrad_phi}
        \begin{split}
        \displaystyle\Delta R_{I}^{\Phi}&=\displaystyle\nabla_{\Phi_J} R_I^{\Phi}\Delta\Phi_J \\
        &= \Aop^{n_\text{el}}_{e=1} \left[\int_{B^e}\left(\frac{N^iN^j}{\Delta t}
    	+ \nabla N^i \cdot \frac{\partial{\boldsymbol{q}}}{\partial{\nabla\Phi}} \cdot \nabla N^j
        - N^i \frac{dF^{\Phi}}{d\Phi} N^j\right)\;\text{d}V\right]\Delta\Phi_J\text{.}
        \end{split}
        \ee
        for any arbitrary increment \(\Delta\Phi_J\). Finally, at each Newton iteration, the field variable at global degree of freedom \(I\) is updated as 
        \be\label{Eq:Update_phi}
        \Phi_I = (\Phi_{I})_{o} - \sum_{J=1}^{n_\text{dof}} \frac{R^{\Phi}_J}{\nabla_{\Phi_J} R_I^{\Phi}}\text{,}
        \ee
        where \((\Phi_I)_0\) is the action potential from the previous Newton iteration.
    
    \subsection{Local Newton iteration for recovery variable}\label{subsec:Localiter_r}
        
        Göktepe and Kuhl\cite{Gktepe2009ElectromechanicsOC} have successfully done a coupled electromechanical analysis of a bi-ventricular generic heart model by considering the recovery variable \(r\) internally, solved using a local Newton iteration as described in Algorithm \ref{algo:localnewton}.
        \begin{algorithm}[h]
        \caption{Finding \(r\) via local Newton iterations}
        \label{algo:localnewton}
        \begin{algorithmic}
            \State{$\phi \gets \phi_n, \quad r \gets r_n$}
            \While{$\lvert R^r \rvert \geq \text{TOL}$}
                \State Compute residual $R^r$
                \State Compute derivative $\displaystyle\frac{\partial R^r}{\partial r}$
                \State Compute increment $\displaystyle\Delta r = -\left[\frac{\partial R^r}{\partial r}\right]^{-1} R^r$
                \State Update $r \gets r_n + \Delta r$
            \EndWhile
            \State Compute $\displaystyle f^\phi, \quad \frac{\text{d}f^\phi}{\text{d}\phi}$
            \State Update history variable $r_n$
        \end{algorithmic}
        \end{algorithm}
        Considering a time interval \([\tau_n,\tau]\) with an increment \(\Delta\tau=\tau-\tau_n\), \autoref{Eq:AP_strong_normed} may be approximated as
        \be\label{Eq:TimeDiscrete_fr}
        \frac{r-r_n}{\Delta \tau}=f^r\text{,}
        \ee
        where \(r\) and \(r_n\) are the values corresponding to the time points \(\tau\) and \(\tau_n\), respectively. The derivative of the residual
        
        \be\label{Eq:Res_r}
        R^r = r-r_n-f^r\Delta \tau
        \ee
        
        is expressed as 
        
        \be\label{Eq:Res_drdr}
        \frac{\partial R^r}{\partial r} = 1 + \left[\gamma + \frac{\mu_1}{\mu_2 + \phi}\left[2r + c\phi\left[\phi-b-1\right]\right]\right]\Delta \tau\text{,}
        \ee
        
        and \(r\) is updated according to Algorithm \ref{algo:localnewton}.

\clearpage
\bibliography{wileyNJD-AMA,Cardiac_PINNs}

\end{document}